\documentclass[english,showpacs,twocolumn,aip,reprint,superscriptaddress]{revtex4-1}

\usepackage{mathptmx}
\usepackage[T1]{fontenc}
\usepackage[latin9]{inputenc}
\usepackage{color}
\usepackage{amsmath}
\usepackage{amssymb}
\usepackage{hyperref}
\usepackage{graphics}
\usepackage{graphicx}
\usepackage{epsfig}
\usepackage{bm}
\usepackage{epstopdf}
\usepackage{mathtools}
\usepackage{soul}
\usepackage{subfigure}   
\usepackage{xmpmulti}
\usepackage{lipsum}

\usepackage{array}
\usepackage{tabularx}
\usepackage{multirow}
\usepackage[thinlines]{easytable}

\def\beq{\begin{equation}}
\def\eeq{\end{equation}}

\def\reff#1{(\ref{#1})}
\def\rhoc{\rho_\mathrm{c}}
\def\rhoi{\rho_\mathrm{i}}

\def\Up{U_\mathrm{p}}

\def\Wcmcm{\mbox{\rm W/cm$^{2}$}}

\def\omegaM{\omega_\mathrm{M}}

\def\omegaC{\Omega_\mathrm{c}}
\def\omegaC0{\Omega_\mathrm{c0}}

\def\vekt#1{\bm{#1}}
\def\vect#1{\vekt{#1}}

\def\pabl#1#2{\frac{\partial #1}{\partial #2}}

\def\v0{v_0}
\def\I0{I_0}

\definecolor{dgreen}{rgb}{0.0, 0.5, 0.0}
\definecolor{dblue}{rgb}{0.0, 0.0, 0.5}
\definecolor{green}{rgb}{0.0,0.5,0.0}

\usepackage{babel}

\begin{document}

\title{Laser-cluster interaction in an external magnetic field: the effect of laser polarization}

\author{Kalyani Swain}
\affiliation{Institute for Plasma Research, Bhat, Gandhinagar, 382428, India}
\affiliation{Homi Bhabha National Institute, Training School Complex, Anushaktinagar, Mumbai 400094, India}


\author{and Mrityunjay Kundu}
\affiliation{Institute for Plasma Research, Bhat, Gandhinagar, 382428, India}
\affiliation{Homi Bhabha National Institute, Training School Complex, Anushaktinagar, Mumbai 400094, India}

\date{\today}
\begin{abstract}
Collisionless absorption of laser energy by an electron via laser-cluster 
interaction in an ambient magnetic field ($B_0$) has recently renewed interest. 
Previously, using a rigid sphere model (RSM) and an extensive 
particle-in-cell (PIC) simulation with linearly polarized (LP) laser light, we 
have shown that an auxiliary field $B_0$ in a transverse direction to 
the laser polarization significantly enhances the laser absorption [Scientific Reports 
{\bf 12},
11256 (2022)]. In this LP case, the
average energy ($\mathcal{E}_A$) of an electron rises near $30-70$ times of 
its 
ponderomotive energy ($U_p$). The two-stage laser absorption by cluster 
electrons has been attributed via anharmonic resonance (AHR) followed by
electron-cyclotron resonance (ECR)
satisfying the improved phase-matching and frequency-matching conditions 
simultaneously.
In the present work, we study the effect of circularly polarized (CP) laser 
fields on the cluster-electron dynamics considering left/right circular 
polarizations with an ambient $B_0$. 
In typical conditions, without $B_0$, we show that both LP and CP light
yield almost the same level of laser absorption (about $3\Up$ or less) by 
an electron. However, with $B_0$, CP light enhances the electron's energy 
{\em further} by $\approx 10-20 \Up$ beyond the previously reported values
$\approx 30-70\Up$ by the LP light. These ejected electrons from cluster 
show narrow cone-like propagation as a weakly relativistic electron beam with 
an 
angular spread $\Delta\theta<5^{\circ}$ and the beam quality improves in CP than 
LP. In all cases, RSM and PIC results show good agreement.


\end{abstract}

\maketitle
\section{Introduction}\label{sec1}
\vspace{-0.25cm}
In the past few decades, table-top size particle accelerators based on
laser-plasma interaction have brought significant progress to replace a
traditional multi-kilometer high-energy particle accelerator. The 
production of a relativistic electron beam (REB) via laser-plasma interaction is 
also of current interest due to its many applications, e.g., in the fast 
ignition technique of inertial confinement fusion and medical fields. Here 
target material plays an important role. As a unique target media, atomic 
nano-clusters with solid-like local density 
may absorb nearly $80\%$ of laser energy compared to bulk solid and gas
targets\cite{Ditmire_PRL78_2732}. Laser-cluster interaction 
(LCI) may produce energetic 
ions\cite{Ditmire_PRL78_2732,Ditmire_PRA57,Ditmire_Nature386,Lezius}, 
neutrals\cite{Rajeev_Nature}, 
electrons\cite{Ditmire_PRA57,Chen_POP_9,Shao_PRL77,Springate_PRA68,Chen_PRE_2002} and 
x-rays\cite{McPherson_Nature370,Chen_PRL104,Jha_2005,Dorchies_Xray,
Kumarappan2001}, {thereby opening avenues for new-generation particle accelerators} and light sources. The fundamental processes involved in LCI 
encompass: (i) inner ionization ---{separation of electrons (and ions) from parent atom, results the generation} of nano-plasma, (ii) outer ionization ---{complete removal of electrons from the cluster boundary}, (iii) coulomb explosion ---{expansion of the ionic background which 
are discussed} in earlier 
works\cite{RosePetruck,Bauer2003,Siedschlag_PRA_67_2003,Snyder_PRL_1996_IG}.

In the collision-less regime of laser 
absorption\cite{Ishikawa,Megi,Jungreuthmayer_2005,Bauer2004} by cluster 
electrons (for laser intensities $\I0>10^{16}\,\Wcmcm$ and wavelength 
$\lambda>600$~nm),
linear resonance (LR) and anharmonic resonance (AHR) are significant. The LR
occurs\cite{Ditmire_PRA53,LastJortner1999,Saalmann2003,Fennel_EPJD_29_2004} for 
a long-duration laser pulse (typically > 50 fs) {when there is} sufficient coulomb explosion of {the cluster (which is over-dense initially; $\rhoi>\rhoc$)}. This happens as the {charge density of ions
$\rhoi(t)$ eventually} decreases and {meets the critical plasma} density $\rhoc = \omega^2/4\pi$, where the \text{dynamic} Mie-plasma frequency $\omegaM(t)=\sqrt{4\pi \rhoi/3}$ {aligns} with the laser frequency $\omega=2\pi c/\lambda$. We use atomic units (a.u.) where $\vert 
e\vert = m_0 =  \hslash = 4\pi\epsilon_0 = 1$ unless noted explicitly. On the other hand,
for short duration laser pulse, LR can not happen since cluster 
is over-dense with $\omegaM(t)>\omega$, and in 
this case AHR plays important role. During AHR, the self-consistent anharmonic potential of the cluster allows the electron oscillation frequency to dynamically align with $\omega$. 
Significance of AHR is discussed frequently {in numerous studies}\cite{MulserPRA,MulserPRL,MKunduPRA2006,MKunduPRL,Kostyukov,Taguchi_PRl,
SagarPOP2016} using {a simple} rigid sphere model (RSM) of {cluster}, particle-in-cell (PIC) and molecular dynamics (MD) simulations.

A comprehensive review for LCI, as given in Table-I of 
Ref.\cite{Swain2022}, reveals 
an interesting observation without an external magnetic field $B_0$.
It shows that, even when the laser has an adequate supply of energy, the 
maximum 
average energy attained by a cluster electron mostly remains close to $\approx 
3.17\Up$ (similar to the 
laser-atom\cite{MorenoEPL_1994,MorenoPRA_1997,MLeinPRL2003} interaction) or 
below. Here, $\Up = I_0/4\omega^2$ is the non-relativistic ponderomotive energy 
of an electron.
In Ref.\cite{Swain2022}, using RSM and a three-dimensional PIC simulation, we 
show that an ambient magnetic field ($B_0$) in transverse orientation to the
linearly polarized laser field of a 5-fs broadband laser pulse (with $\lambda = 
800$~nm and intensities $\I0 \approx 10^{16} - 
10^{18}\,\Wcmcm$) can significantly enhance average energy of an electron 
${\mathcal{E}}_A$ up to 
$\approx 36-70\,\Up$ (more than $12-36$ fold) while interacting with a deuterium 
cluster. This {substantial improvement in laser energy absorption is a two stage process}: with AHR as the 
{\em first stage} and electron cyclotron resonance (ECR) or relativistic ECR 
(RECR) as the {\em second stage}, where electron cyclotron frequency 
$\Omega_\mathrm{c}=\vert e \vert B_0/m_0\gamma = \omegaC0/\gamma$ meets the 
laser frequency $\omega$ and simultaneously satisfies the phase matching 
condition. In a subsequent work \cite{Kalyani_PRA_2023} we increase 
the cluster size ($R_0\approx 2.2 \rightarrow 4.4$~nm) and show that, under 
similar conditions, per electron energy ${\mathcal{E}}_A$ remains 
almost the same ($\approx 36-70\,\Up$) near ECR, but {in 
the regime of $100\%$ outer-ionization (at high intensities)} the
net absorption increases almost linearly with the number of electrons $(N)$.
We find that laser-coupled cluster electrons form a nearly mono-energetic weakly 
relativistic, spirally narrow beam that traverses a few hundred of $R_0$ (or on 
the order of $\lambda$) with momentum $p/m_0c> 1.7$ in an ambient
magnetic field near ECR, which may not be possible only with the laser field. 
Also, as the cluster size increases, the intensity of electron beam increases 
with a greater number of energetic electrons at a restricted angle of $\theta_r 
\approx 7^\circ - 10^\circ$ (in the position space) and $\theta_p \approx 
58^\circ - 62^\circ$ (in the momentum space) w.r.t. the laser propagation direction $\vekt{z}$ for 
$I_0 = 7.13\times10^{16} - 1.83\times10^{17} \Wcmcm$.

\begin{figure}[]
	\includegraphics[width=0.4\textwidth]{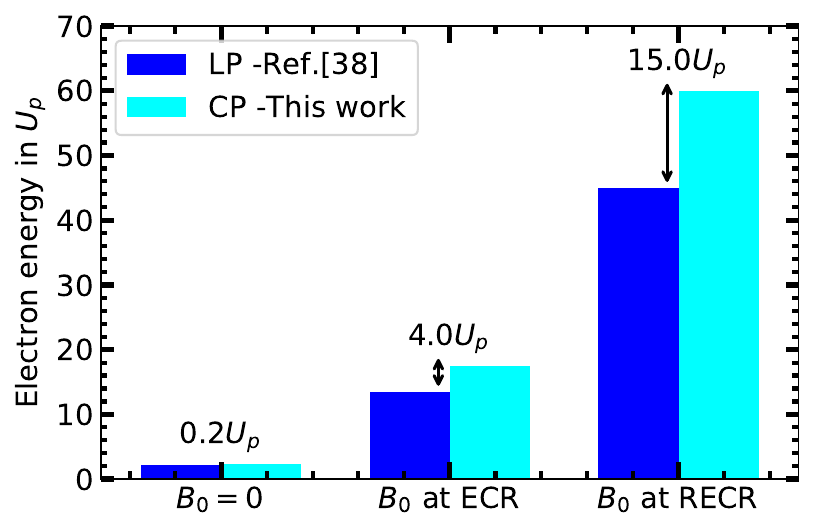}
	\vspace{-0.25cm}
	\caption{Schematic of comparison of average energy gain by a laser-driven cluster-electron for LP and CP fields with/without $B_{ext}$.}
	
	\vspace{-0.5cm}
	\label{Fig1}
\end{figure}

In this paper we study the effect of circular polarized (CP) laser field on 
the energy absorption and subsequent dynamics of cluster electrons in presence 
of an ambient magnetic field $\vekt{B}_{ext} =\pm B_0\vekt{\hat{z}}$ using RSM
and PIC simulations. We show that, for CP laser fields absorbed energy 
per electron ${\mathcal{E}}_A$ is further enhanced and it is 
$10-20~\Up$ more compared to the linearly polarized (LP) cases\cite{Swain2022,Kalyani_PRA_2023}. 
To provide a quick summary, in Figure~\ref{Fig1} we show a schematic of comparison of average energy gain by a laser-driven cluster-electron for LP and CP fields with/without $B_{ext}$ for $I_0 = 1.83\times 10^{17} \,W/cm^2$.
Additionally, we find that 
right circular polarized (RCP) and left circular polarized (LCP) laser fields 
yield the same amount of absorption, but the direction of $\vekt{B}_{ext}$ for 
LCP (RCP)
should be along $-\hat{z}$ ($+\hat{z}$) w.r.t. laser propagation 
direction. Note that the
field components of a CP light does not vanish simultaneously, while for LP 
it can reach zero. The phase 
matching condition show that one of the field components with CP light may 
maintain the required phase with non-zero amplitude. This leads to higher 
absorption of the CP laser pulse than the LP case in presence of 
$\vekt{B}_{ext}$. In addition, we find that the quality of the emitted electron 
beam in terms of divergence and energy is improved in CP than LP.

In Sec.\ref{secPulse} laser pulse and cluster parameters are introduced. In 
Sec.\ref{sec2} we describe RSM and PIC method followed by the 
simulation results. In Sec.\ref{sec6} we summarize this work.

\vspace{-0.25cm}
\section{Laser pulse and cluster parameters}\label{secPulse}
\vspace{-0.25cm}
We assume a laser pulse \cite{MKunduPRA2012,SagarPRA2018,Swain2022} 
propagating in $\vekt{z}$ with a vector potential
\begin{eqnarray}\label{eq:vectorpotential}
\vekt{A}(t') =  \frac{E_0}{\omega} \sin ^2 \left(\frac{\omega t'}{2n}\right)\!\left[\delta \cos (\omega t') \hat{\vekt{x}} +\! \sqrt{1\!-\!\delta^2} \sin (\omega t') \hat{\vekt{y}}\right]\!\!.
\end{eqnarray}
Where $t' = t - {z}/c$ is the retarded time, $\tau = n T$ is the pulse duration such that $0 \leq t' \leq \tau$, and $n=$ is the number of laser period $T$. The factor $\delta = 1$ is for LP with polarization in $\vekt{x}$, and $\delta = \pm 1/\sqrt{2}$ are for left/right CP (LCP/RCP) respectively. The field strength is $E_0=\sqrt{I_0}$.
%
Laser electric and magnetic fields $\vekt{E}_l$, $\vekt{B}_l$ are obtained from
\begin{equation}\label{eq:laserfieldE}
%
\vekt{E}_l (t')  = -\pabl{\vekt{A}}{t}, \,\,\,\,\,\,\,\,
\vekt{B}_l (t')  = \hat{\vekt{z}} \times \vekt{E}_l (t') / c.
\end{equation}

\noindent
The laser pulses \reff{eq:vectorpotential} show broad-band nature (for short pulses) due to three
discrete frequencies $\omega_1 = \omega, \omega_2 = (1+1/n)\omega$, $\omega_3 
= (1-1/n)\omega$. We assume $\lambda = 800$~nm,
$n=5$, and $\tau = nT\approx 13.5$~fs ($\tau_{fwhm}\sim 5$~fs) in this work.

A deuterium cluster of number of atoms $N = 2176$ is chosen. With the
Wigner-Seitz radius $r_w\approx 0.17$~nm, cluster radius is $R_0=r_w 
N^{1/3} \approx 2.2$~nm. Since $R_0\ll\lambda$, the dipole approximation 
$z/\lambda \ll 1$ may be assumed. {For $\lambda = 800$~nm laser, the critical density is $\rhoc \approx 1.75\times 10^{27} m^{-3}$. Hence,} the cluster is $\rho_i/\rhoc \approx 27.1$ times overdense with $(\omegaM/\omega)^2 \approx 9.1$.

\vspace{-0.25cm}
\section{Model and numerical simulations}\label{sec2}
\vspace{-0.25cm}
We pursue two numerical approaches in this work, namely a simple RSM 
and self-consistent PIC simulation. These are already described in earlier 
works \cite{MKunduPRA2006,MKunduPRL,Swain2022,Kalyani_PRA_2023}. Therefore, 
we describe them in short to maintain the readability.

\vspace{-0.25cm}
\subsection{Rigid sphere model (RSM)}\label{subsec1}
\vspace{-0.25cm}
The rigid sphere model (RSM) simplifies the description of the dynamics of an 
electron in the case of laser-cluster interaction. It also provides a useful 
framework for understanding many physics aspects of laser absorption. In
this model, cluster is considered as a pre-ionized nano-plasma {which is spherical in
shape}. For simplicity, {the ionic background is considered to be of the same
as that of} the initial cluster radius $R_0$. {As we focus on the short-pulse laser duration, the ions are {treated as} stationary (or frozen) and their dynamics are omitted} in the RSM.
The ionic charge density $\rhoi =3 N 
Z/4\pi R_0^3$ defines the Mie-plasma frequency $\omegaM^2 = 4\pi\rhoi/3$ for
fully ionized number of atoms $N$ each of charge $Z=1$. RSM is widely discussed 
in earlier 
works\cite{SagarPOP2016,SagarPRA2018,MulserPRA,MulserPRL,MKunduPRA2006,MKunduPRL,Krishnan_PRL,Krishnan_2014,Swain2022} which include the dynamics of a 
single electron as well as non-interacting multi-electrons to mimic a more 
realistic scenario.
The potential $\Phi_I(r)$ and the space-charge field $\vekt{E}_{sc} (\vekt{r})$ in RSM are given by
\vskip -0.25cm
\noindent
\begin{minipage}{.48\linewidth}
	\begin{equation}\nonumber  
		\phi_I (r)\! =\! \omegaM^2 R_0^2
		\begin{cases}
			{3}/{2}\!-\!{r^2}/{2 R_0^2} \\
			{R_0}/{r}
		\end{cases}
	\end{equation}
\end{minipage}%
\begin{minipage}{.49\linewidth}
	\begin{equation}\label{eom1aRSM} 
		\vspace{-0.4cm}
		\!\!\!\!\!;~\vekt{E}_{sc} (\vekt{r})\!=\!
		\begin{cases}
			\omegaM^{2} \vekt{r}  \\
			\omegaM^{2} R_0^3 \vekt{r}/{r^3}
		\end{cases}
	\end{equation}
\end{minipage}
\noindent 
\\\\\\
The top and bottom lines are for $r\le R_0$ and $r>R_0$ respectively. An electron interacts with the applied fields ($\vekt{E}_l, \vekt{B}_l$), and $\vekt{B}_{ext}$ {including the coulomb field $E_{sc}(r)$} obeying the following equations
%
%
\begin{flalign}\label{eom1bRSM}
	\frac{d\vekt{p}}{dt} & =  q\left[ \left(\vekt{E}_l + \vekt{E}_{sc}(\vekt{r}) \right) + \vekt{v}\times \left(\vekt{B}_l + \vekt{B}_{ext} \right)\right] \\ \label{eom1cRSM}
	\frac{d\vekt{r}}{dt} & = \vekt{v} = \frac{\vekt{p}}{\gamma m_0}
	\\ \label{eom1dRSM}
	\frac{d({\gamma} m_{0} c^2) }{dt} & = q {\vekt{v}}.\left(\vekt{E}_l + \vekt{E}_{sc}(\vekt{r}) \right)
\end{flalign}
%
where $\gamma = 1/\sqrt{1-v^2/c^2}= \sqrt{1+p^2/m_{0}^2 c^2}$ is the 
relativistic $\gamma$-factor for the electron. $m_{0}, q, \vekt{r}, \vekt{v}, 
\vekt{p}$ are {the rest-mass ($m_0 = 1$ in a.u.), charge ($q = e = -1$ in a.u.),} position, velocity and linear momentum respectively. In earlier 
works\cite{Swain2022,Kalyani_PRA_2023}, we have demonstrated that the average 
energy absorption of multiple non-interacting electrons in the RSM in an 
ambient magnetic field is very similar to that of a single electron in RSM. 
Hence, we do not reiterate those RSM results with multiple non-interacting 
electrons here. Instead, we consider a 
single electron in the RSM when exposed to a CP light and compare the RSM 
results with the self-consistent PIC results.

%
\subsubsection{Laser absorption by an electron with LP and CP laser
field}\label{subsec1}
\vspace{-0.25cm}
In the previous works \cite{Swain2022, Kalyani_PRA_2023} we studied 
interaction of LP laser pulses with deuterium clusters 
in presence of ambient magnetic fields 
in the range of $\vert\vekt{B}_{ext}\vert = \vert B_0\vekt{\hat{z}}\vert = 0-20$~kT. 
The intensity of LP pulses were $I_0 \approx 7\times10^{16}-2\times10^{17} 
W/cm^2$ with fixed wavelength $800$~nm and pulse duration of 
$\sim13$fs. The average energy absorption by an 
electron was enhanced in the range of $\mathcal{E}_A\approx 35-70 U_{p}$ due to ECR (where, $\Omega_\mathrm{c} = \omegaC0 = \omega$) for $\vert\vekt{B}_{ext}\vert = \vert B_0\vekt{\hat{z}}\vert = 0-20$~kT from a value of $\mathcal{E}_A\approx 2 U_{p}$ (or below) without $\vekt{B}_{ext}$.
Note that, in those cases $\vekt{B}_{ext} = B_0\vekt{\hat{z}}$ were chosen along the laser propagation
$+\vekt{\hat{z}}$. However, the dynamics of cluster electrons for a reversed $\vekt{B}_{ext} =
-B_0\vekt{\hat{z}}$ is still not addressed for the LP light. The effect of CP 
laser fields with $\vekt{B}_{ext} =
\pm B_0\vekt{\hat{z}}$ and corresponding ECR absorption have not been reported 
so far for laser-cluster interaction.

\begin{figure}[]
\includegraphics[width=0.45\textwidth]{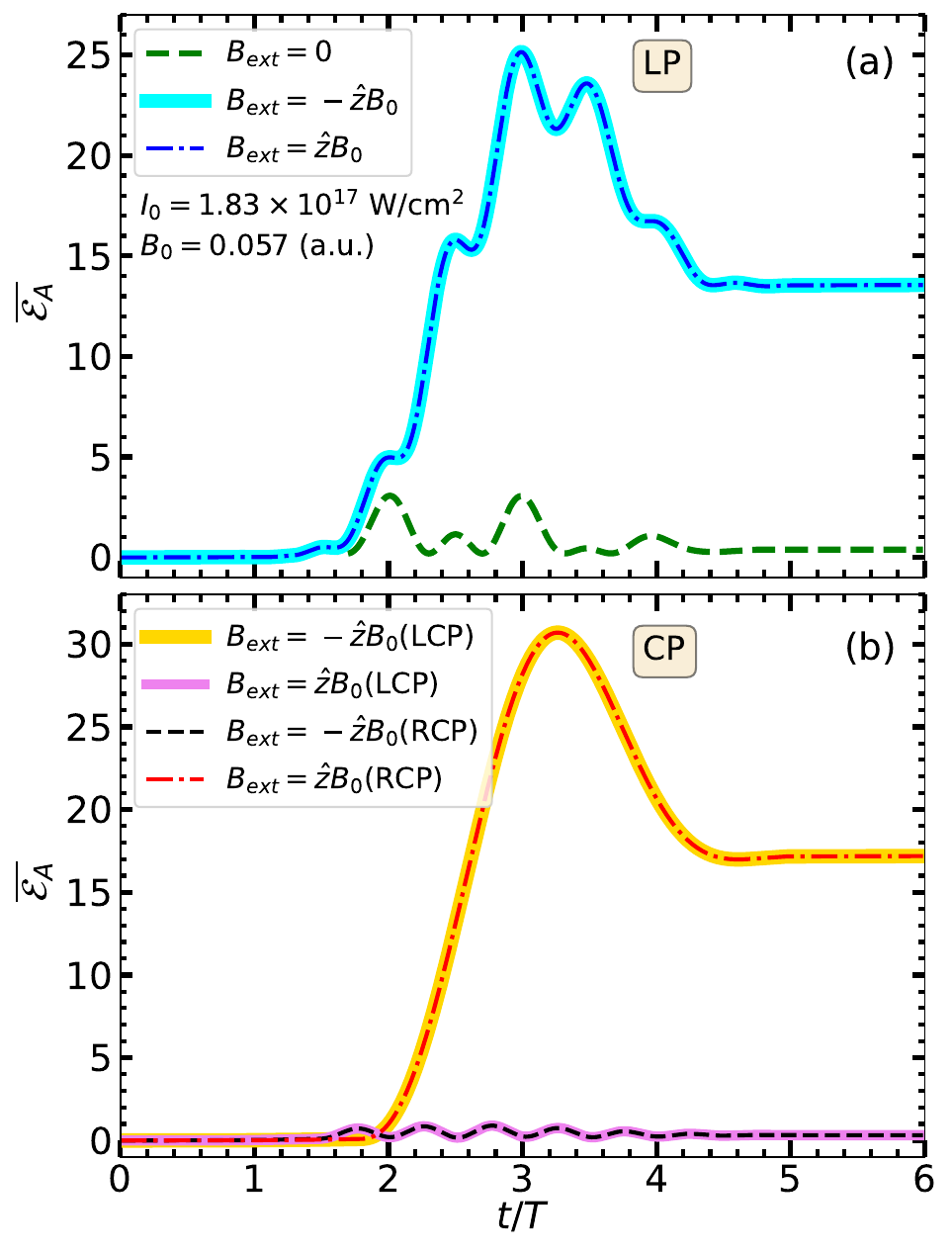}
	\vspace{-0.25cm}
	\caption{Time vs total laser energy absorption $\overline{\mathcal{E}_A}$ in $\Up$ for a single electron in RSM for (a) LP laser when $\vekt{B}_{ext} = \pm B_0\vekt{\hat{z}}$ (blue dash-dot and cyan solid lines; black dash-dot and light grey solid lines in grey scale), and $\vekt{B}_{ext} = 0$ (green dashed line; dark grey dashed lines in grey scale), (b) RCP laser field when $\vekt{B}_{ext} = \pm B_0\vekt{\hat{z}}$ (dash-dot red, dashed black lines; dark grey dash-dot and black dashed lines in grey scale), and LCP laser field when $\vekt{B}_{ext} = \pm B_0\vekt{\hat{z}}$ (purple and yellow solid lines; dark grey and light grey solid lines in grey scale). Here $B_0 =  0.057$a.u.= 13.37 kT and intensity $I_0 = 1.83\times 10^{17} W/cm^2$.}
	\label{Fig2}
\end{figure}

In Figure~\ref{Fig2} we compare the temporal variation of normalized absorbed energy
($\overline{\mathcal{E}_A} =
{\mathcal{E}}_A/\Up$) by a single electron in the RSM with LP (Fig.\ref{Fig2}a) 
and CP (Fig.\ref{Fig2}b)
laser fields for two opposite orientations of $\vekt{B}_{ext} = \pm 
B_0\vekt{\hat{z}}$. The CP cases include left/right circular polarizations 
(LCP/RCP). The laser-pulse is $n=5$-cycle as in Eq.\reff{eq:vectorpotential} with intensity $I_0 
= 1.83\times 10^{17}\, W/cm^2$. The electron is initially placed at the center 
of the cluster with zero velocity. When no external magnetic field
($B_0 = 0$), energy absorption by the electron is $\overline{\mathcal{E}_A} 
\approx 0.5$ for all cases of LP and CP, in the end of the 
pulses. However, in the presence of an ambient magnetic field at the ECR value 
of 
$B_0 = \omega=0.057$~a.u. ($\approx 13$ kilo Tesla) with LP laser 
we obtain
$\overline{\mathcal{E}_A}\sim 13$ for $\vekt{B}_{ext} = \pm B_0\vekt{\hat{z}}$ 
(solid and dashed-dot, Fig.\ref{Fig2}a). 
It means that the value of absorption $\overline{\mathcal{E}_A}$ is unaltered 
irrespective of the orientation of $\vekt{B}_{ext} = \pm B_0\vekt{\hat{z}}$ for 
the LP light.

At the ECR value of $B_0 = \omega=0.057$~a.u., when
$\vekt{B}_{ext} = + B_0\vekt{\hat{z}}$, the 
energy $\overline{\mathcal{E}_A}$ is enhanced up to $\approx 17$ for RCP 
(dashed-dot); and
for $\vekt{B}_{ext} = -B_0\vekt{\hat{z}}$, absorption 
drops to $\overline{\mathcal{E}_A} \approx 0.5$ as without 
$\vekt{B}_{ext}$ for RCP (Fig.\ref{Fig2}b). For LCP with $\vekt{B}_{ext} = 
-B_{0}\vekt{\hat{z}}$ we obtain equal amount of enhancement in 
$\overline{\mathcal{E}_A}\approx 17$ (solid, light-grey) similar to the RCP with
$\vekt{B}_{ext} = + 
B_0\vekt{\hat{z}}$. Again, when $\vekt{B}_{ext} = + 
B_0\vekt{\hat{z}}$ for LCP laser, the 
absorption drops to $\overline{\mathcal{E}_A} \approx 0.5$ as without 
$\vekt{B}_{ext}$. Clearly, laser absorption depends on the orientation of 
$\vekt{B}_{ext}$ for RCP/LCP laser fields. In short, RCP favours $\vekt{B}_{ext} 
= + B_0\vekt{\hat{z}}$ and LCP favours $\vekt{B}_{ext} = - B_0\vekt{\hat{z}}$
for the electron's energy enhancement due to the co-rotating cyclotron motions 
with these respective CP fields. Importantly, the value of 
$\overline{\mathcal{E}_A}$ is more than $4\,\Up$ higher in the 
end of the laser pulse, for both RCP/LCP fields than the
respective LP cases at the ECR.
 
\begin{figure*}[]
	\centering
 	\includegraphics[width=0.95\textwidth]{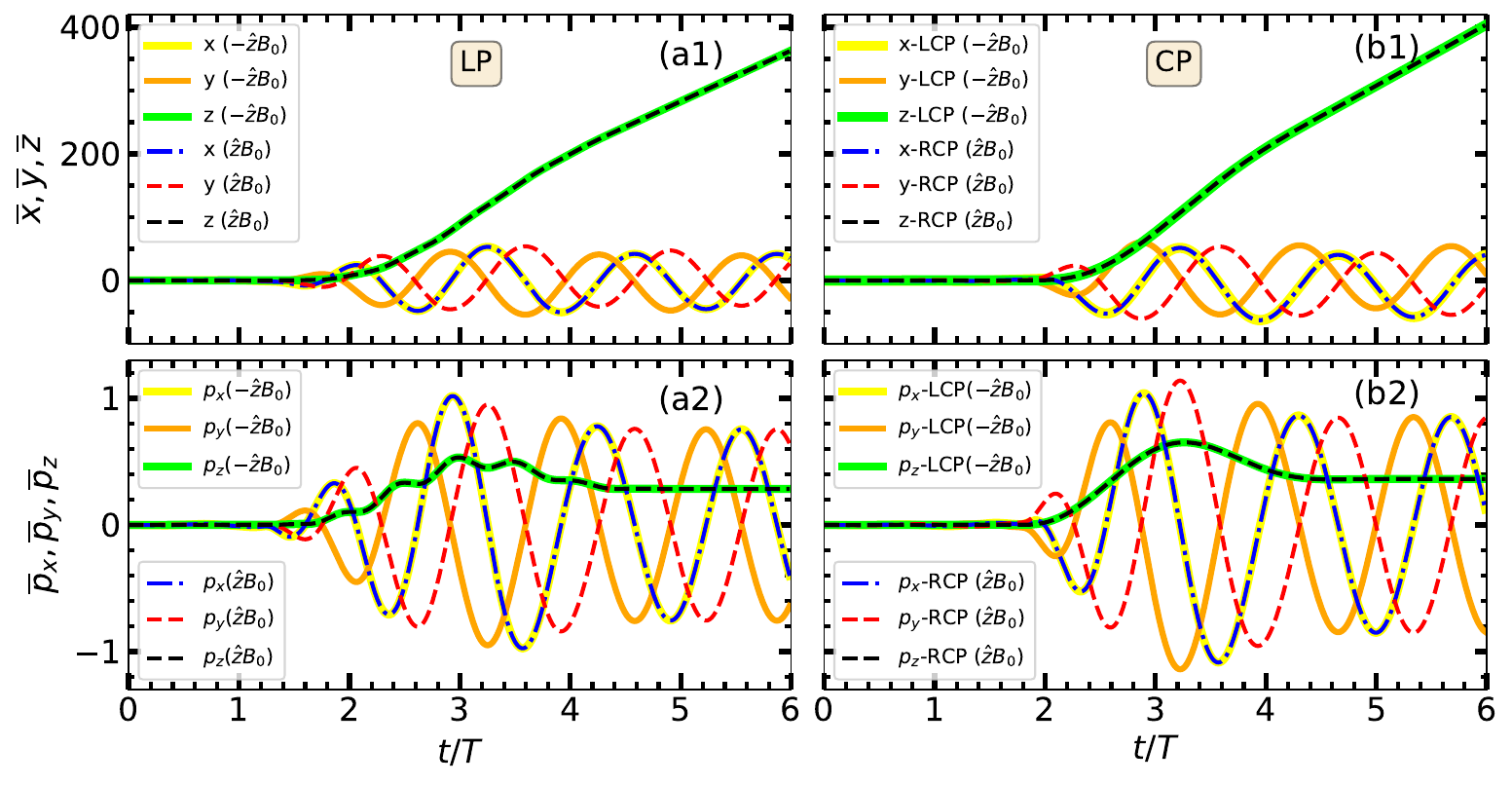}
	\caption{Time vs normalized position $\overline{x}, \overline{y}, \overline{z}$ and normalized momentum $\overline{p}_x, \overline{p}_y, \overline{p}_z$ of the RSM electron for LP (a1,a2) and CP (b1,b2) laser fields when $B_{ext} = \pm \hat{z}B_0$. These trajectories correspond to those enhanced absorption cases of $\overline{\mathcal{E}_A} \approx 13$ (LP case, blue, cyan lines in Fig.\ref{Fig2}a) and $\overline{\mathcal{E}_A} \approx 17$ (CP case, yellow and red lines in Fig.\ref{Fig2}b) for $I_0 = 1.83\times 10^{17} W/cm^2$ and $B_0=0.057$ a.u. at ECR.}
	%
	%
	\label{Fig3}
\end{figure*}

\vspace{-0.5cm} 
In the case of LP laser, $\overline{\mathcal{E}_A}$ is shown (Fig. \ref{Fig2}) 
unaltered irrespective of the orientation $\vekt{B}_{ext} = \pm 
B_0\vekt{\hat{z}}$, but the corresponding electron dynamics may 
differ as shown in Fig.\ref{Fig3}. The electron dynamics in $x, z$ do not 
change for $\vekt{B}_{ext} = \pm B_0\vekt{\hat{z}}$ in the position and the 
momentum space (Fig.\ref{Fig3}a1, \ref{Fig3}a2). However, the dynamics in 
y-direction in both position and in momentum space are in opposite phase for 
$\vekt{B}_{ext} = \pm B_0\vekt{\hat{z}}$.
It shows clock-wise rotation for $\vekt{B}_{ext} = + B_0\vekt{\hat{z}}$ and anticlock-wise rotation for $\vekt{B}_{ext} = - B_0\vekt{\hat{z}}$ in the x-y plane, but the electron always moves forward in $\vekt{\hat{z}}$.

The trajectory of the electron in the position and momentum space 
(Fig.~\ref{Fig3}b1,\ref{Fig3}b2) for LCP field with 
$\vekt{B}_{ext} = - B_0\vekt{\hat{z}}$ resembles with the LP case when 
$\vekt{B}_{ext} = - B_0\vekt{\hat{z}}$ and the RCP case 
resembles with the LP case when $\vekt{B}_{ext} = + 
B_0\vekt{\hat{z}}$. However, the amplitude of respective 
trajectories are different for LP and CP fields, particularly $z,p_z$ 
are higher for CP than LP. The higher $p_z$ explains higher energy for the CP 
case (Fig.\ref{Fig2}b) than the LP case (Fig.\ref{Fig2}a).

\vspace{-0.25cm}

%
%
%
\subsubsection{Phase dynamics of an electron with LP and CP laser field}\label{subsec4}
\vspace{-0.25cm}
To comprehend the enhancement of absorbed energy, it is important to analyze the 
temporal phase-angle dynamics between the velocity of the electron and the 
associated electric field. It is clear from Eq.\reff{eom1dRSM} that the phase 
angle $\Delta \psi$ between the electron velocity ($\vekt{v}$) and the driving 
electric field ($\vekt{E}$) plays crucial role to determine the rate of 
absorption ${d({\gamma} m_{0} c^2) }/{dt}$. When the phase angle $\Delta \psi$ 
is an odd integral multiple of $\pi/2$, the rate of absorption is ${d({\gamma} 
m_{0} c^2) }/{dt} = 0$. However, the absorption rate becomes non-zero as $\Delta 
\psi$ deviates from $\pi/2$ (or odd integral multiple of $\pi/2$) and it reaches 
maximum near $\pi$ (or integral multiple of $\pi$). In the case of the LP light, 
polarized along $x$-direction, the x-component of the field is dominant and 
${d({\gamma} m_{0} c^2) }/{dt} = q v_x E_x \approx q v_x E_l$ holds outside the 
cluster where space-charge field is negligible.
In our earlier work~\cite{Swain2022}, by numerically retrieving the phase angles $\psi_{v_x}$, $\psi_{E_x}$ of respective $v_x, E_x$ and the phase difference $\Delta \psi = \psi_{v_x}-\psi_{E_x}$ for a LP laser field, we demonstrated the importance of an ambient magnetic field $\vekt{B}_{ext} = +B_0\vekt{\hat{z}}$ in the phase dynamics of an electron.
Equation~\reff{eom1dRSM} indicates that, though $B_{ext}$ does not have any 
direct impact on the energy absorption, it may re-orient the angle $\psi_{v_x}$ 
and allow $\Delta\psi$ to remain near $\pi$ (or integral multiple of $\pi$) for 
a prolonged duration which lead to a significant enhancement in energy 
absorption at ECR.

\begin{figure}[]
	\includegraphics[width=0.47\textwidth]{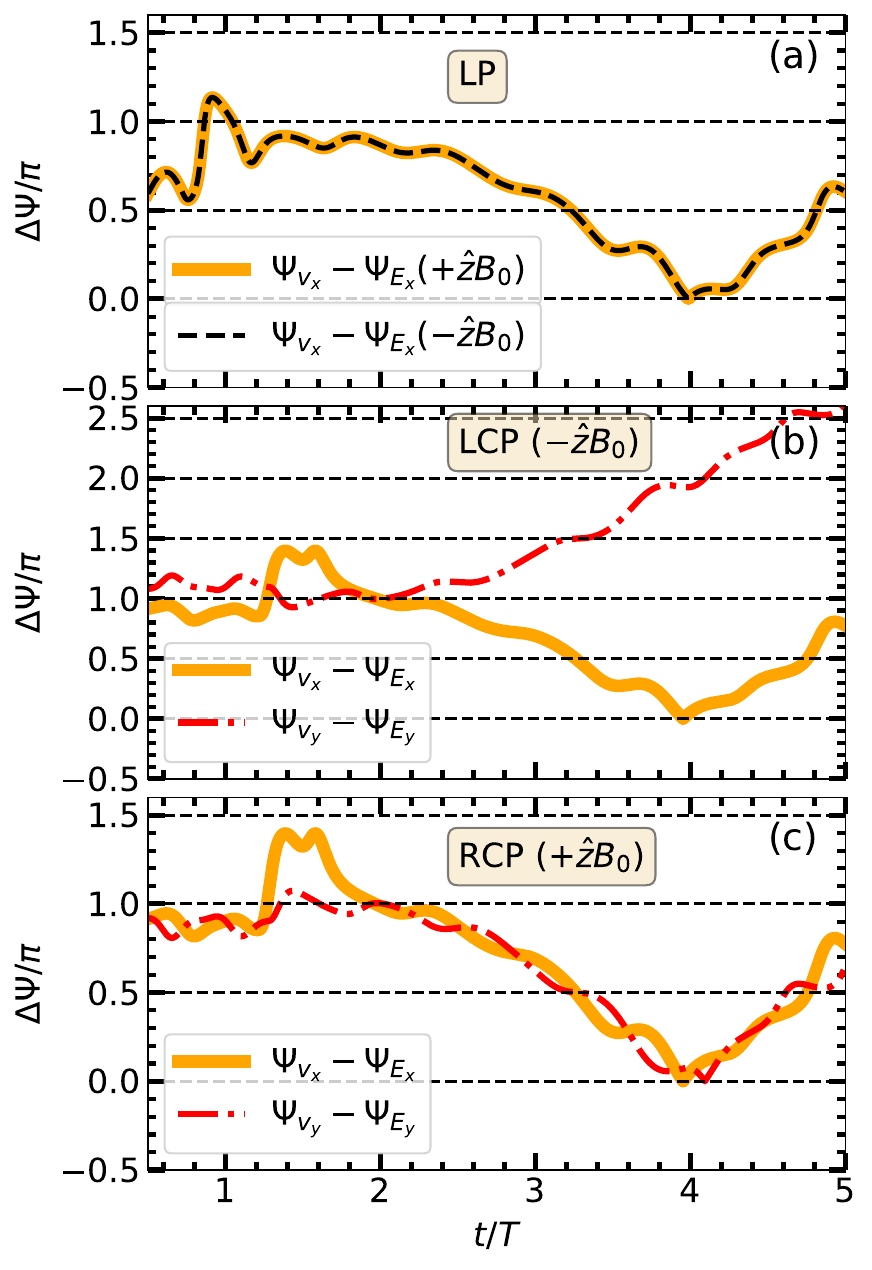}
	\vspace{-0.35cm} 
	\caption{Phase dynamics $\Delta \psi = \psi_{v_x}-\psi_{E_x}$ (for $x$-motion) and $\Delta \psi = \psi_{v_y}-\psi_{E_y}$ (for $y$-motion) versus time are shown for the RSM electron in Fig.\ref{Fig3}. For LP case (a) with $B_{ext} = -\hat{z}B_0$ (dash-dot lines) and $B_{ext} = \hat{z}B_0$ (solid lines), $\psi_{v_x}-\psi_{E_x}$ show exact match. For LCP field (c) with $B_{ext} = -\hat{z}B_0$ and RCP field (b) with $B_{ext} = \hat{z}B_0$, $\psi_{v_x}-\psi_{E_x}$ show a good match (solid lines in b,c); but $\psi_{v_y}-\psi_{E_y}$ are different (dash-dot lines in b,c). All parameters are same as in Fig.\ref{Fig3}.}
	
	\vspace{-0.25cm}
	\label{Fig4}
\end{figure}

The phase-angle dynamics with LP laser field when $\vekt{B}_{ext} = -B_0\vekt{\hat{z}}$ as well as CP laser fields with $\vekt{B}_{ext} = \pm B_0\vekt{\hat{z}}$ have not been discussed yet. 
In Fig.~\ref{Fig4}a, we plot $\Delta \psi = \psi_{v_x}-\psi_{E_x}$ for the LP 
laser field with $\vekt{B}_{ext} = \pm B_0\vekt{\hat{z}}$ corresponding to the 
energy absorption curves (solid and dashed-dot) in Fig.~\ref{Fig2}a. In both 
the cases, $\Delta \psi$ are maintained near $\pi$ for the laser pulse duration 
$t/T \approx 1-2.5$ that lead to prominent increase in absorption in 
Fig.\ref{Fig2}a around this time. Also, the exact superposition of $\Delta 
\psi$ values shows that the phase-angle dynamics for a LP laser field is 
unaltered irrespective of the direction of $\vekt{B}_{ext} = \pm 
B_0\vekt{\hat{z}}$. As a result, the energy absorption is also unaltered in 
Fig.\ref{Fig2}a irrespective of the direction of $\vekt{B}_{ext} = \pm 
B_0\vekt{\hat{z}}$.

In the case of LP laser field, $\Delta \psi= 
\psi_{v_x}-\psi_{E_x}$ is calculated only for the laser polarization direction 
$x$ in Fig.\ref{Fig4}a. However,
for the CP case two different values of $\Delta \psi = \psi_{v_x}-\psi_{E_x}$ and $\Delta \psi = \psi_{v_y}-\psi_{E_y}$ are calculated due to two components of the driving CP fields.
Hence, in Figs.~\ref{Fig4}b and \ref{Fig4}c we show $\psi_{v_x}-\psi_{E_x}$ and 
$\psi_{v_y}-\psi_{E_y}$ for the LCP laser with $\vekt{B}_{ext} = 
-B_0\vekt{\hat{z}}$ and for RCP laser with $\vekt{B}_{ext} = +B_0\vekt{\hat{z}}$ 
corresponding to the ${\overline{\mathcal{E}}_A}$ curves (solid light-grey and dashed-dot)
in Fig.~\ref{Fig2}b respectively.
In both cases, LCP with $\vekt{B}_{ext} = -B_0\vekt{\hat{z}}$ and RCP with 
$\vekt{B}_{ext} = +B_0\vekt{\hat{z}}$, corresponding to 
${\overline{\mathcal{E}}_A}\sim 17$ in Fig.\ref{Fig2}b, we find $\Delta \psi = 
\psi_{v_x}-\psi_{E_x}$ for the $x$-motion are exactly similar (solid lines) as
shown in Figs.~\ref{Fig4}b and \ref{Fig4}c.
The dynamics of $\Delta \psi = \psi_{v_y}-\psi_{E_y}$ for the $y$-motion are,
however, different for both the cases (dashed lines in
Figs.~\ref{Fig4}b,\ref{Fig4}c). One can identify that, when $\Delta \psi = 
\psi_{v_x}-\psi_{E_x}$ deviates from $\pi$ near $t/T=1.5$, the other $\Delta 
\psi = \psi_{v_y}-\psi_{E_y}$ still stays close to $\pi$. Until $t/T \sim 2.8$
of the 5-cycle pulse, either one of the phase-angles or both remain close to 
$\pi$.
Clearly, the combined effect of $\psi_{v_x}-\psi_{E_x}$ and 
$\psi_{v_y}-\psi_{E_y}$ for the CP lights allows net $\Delta \psi$ to stay near 
$\pi$ for a prolonged duration $\approx 2.8$-cycle from the beginning of the laser
pulse. This improved phase matching in the case of CP fields lead to the {\em 
further} enhancement of ${\mathcal{E}}_A$ by $\sim 4 \Up$ compared to
the LP laser field in Fig.\ref{Fig2} in the end of the pulses. In the next 
section it is shown that ${\mathcal{E}}_A$ may enhance further by $\sim 
10-20 \Up$ with CP laser.

\subsection{Particle-in-cell (PIC) simulation results}\label{subsec2}
\vspace{-0.25cm}
The results from RSM as described above in Sec.\ref{subsec1} gives a fundamental
understanding of the physical process responsible for the higher absorption of 
laser energy by a cluster electron in the CP light. However, particle 
interactions can not be taken into account in this single-particle picture of 
the RSM. To treat the all kinds of particle-interactions starting from the
creation of the nano-plasma self-consistently, we now use 3D-PIC 
simulation\cite{MKunduPRL,MKunduPRA2006,MKunduPRA2007,Popruzhenko2008,
MKunduPOP2008,MKunduPRA2012,MKundu_Thesis} as in the previous 
work~\cite{Swain2022}. The PIC code was upgraded for a hybrid 
method~\cite{Kalyani_PRA_2023}, particularly to treat the electrons outside the 
computational box with MD simulations\cite{SagarPOP2016,SagarPRA2018} to study 
the effect of ambient magnetic field on the energy absorption by the cluster 
electrons and propagation of these energetic electrons as a conical-spiral beam~\cite{Kalyani_PRA_2023}. This upgraded PIC code is used for the
present study where a deuterium cluster with $R_0 = 2.2$~nm and number of atoms 
$N=2176$ is placed at the center of the simulation box.
When the laser field $\vekt{E}_l(t)$ interacts with the neutral deuterium atoms, the ionization from D to D$^{+}$ happens via the over-the-barrier ionization (OBI\cite{HBethe}) process as it reaches a critical field strength $E_c = \vert \vekt{E}_l(t)\vert = I_p^2(Z)/4 Z$ ({where, $I_p(Z)$ stands for} the ionization potential). Subsequently, the electrons and ions generated though OBI creates the {coulomb field (or space charge field)} $\vekt{E}_{sc}(\vekt{r},t) = -\vekt{\nabla}\phi(\vekt{r},t)$, where $\phi(\vekt{r},t)$ is the time-dependent potential that starts from zero. More details of the PIC code are given in Refs.\cite{MKunduPRL,MKunduPRA2006,MKunduPRA2007,Popruzhenko2008,MKunduPOP2008,MKunduPRA2012,MKundu_Thesis,Swain2022,Kalyani_PRA_2023}

A PIC ion/electron have {the charge to mass ratio same} as that of an {actual} ion/electron. Governing equation of motion in PIC simulation for the $j$-th electron and $k$-th ion is 
\begin{align}\label{eom1a}
\frac{d\vekt{p}_{j\vert k}}{dt}\! &=  q_{j\vert k}\! \left[ \left(\vekt{E}_l(t) + \vekt{E}_{sc}(\vekt{r}_{j\vert k},t) \right) + \vekt{v}_{j\vert k}\times \left(\vekt{B}_l + \vekt{B}_{ext} \right)\right]
\\ \label{eom1b}
\frac{d\vekt{r}_{j\vert k}}{dt} & = \vekt{v}_{j\vert k} = \frac{\vekt{p}_{j\vert k}}{\gamma_{j\vert k} m_{j\vert k}}
\end{align}
%
%
\noindent
where $\vekt{p}_{j \vert k} = m_{j\vert k} \vekt{v}_{j\vert k} \gamma_{j\vert k}, \vekt{v}_{j\vert k}, \vekt{r}_{j \vert k}, m_{j\vert k}, q_{j\vert k}, \gamma_{j\vert k}$ represents momentum, velocity, position, mass, and charge of a PIC electron/ion respectively. Here, $\gamma_{j\vert k} = \sqrt{1+p_{j\vert k}^2/m_{j\vert k}^2 c^2}$ is the relativistic factor. In this case, $m_j = m_0 = 1$, $m_k = M_0 = 2\times 1836$, $q_j = -1$ and $q_k = 1$ in a.u.. Initially, the charge density $\rho_G$ is calculated and the Poisson's equation $\nabla^2\phi_G = -\rho_G$ is solved on the numerical grids using time-dependent monopole boundary condition (subscript $G$ represents grid values of {charge density and potential}). We then compute the corresponding potential $\phi(\vekt{r}_{j\vert k},t)$ by interpolating $\phi_G$ to the particle {location}. Further, by differentiating the interpolated $\phi(\vekt{r}_{j\vert k})$ analytically\cite{MKundu_Thesis}, the electric field $\vekt{E}_{sc}(\vekt{r}_{j\vert k}) = -\vekt{\nabla} \phi(\vekt{r}_{j\vert k})$ is calculated at $\vekt{r}_{j\vert k}$.
The same laser fields \reff{eq:laserfieldE} are employed in PIC and the same velocity verlet method (VVM) is used to solve the equations~\reff{eom1a}-\reff{eom1b} as in RSM.
Since, the laser field strength is high, we consider the collisionless interaction among electrons and ions.
The sum of kinetic energy $KE = \sum_l p_l^2/2 m_l$ and potential energy $PE = \sum_l q_l\phi_l$ of all electrons and ions leads to the total energy $\mathcal{E}(t) = \sum_l q_l\phi_l + p_l^2/2 m_l$ at time $t$.
We also calculate the final average absorbed {laser} energy $ {\mathcal{E}_A} = 
\mathcal{E}(\tau)/N$ per electron in the end of {the pulse} at $\tau=nT$ 
with $n=5$-cycle.
In the PIC simulation, it important to choose the numerical parameters (e.g. spatial and temporal resolution, grid size, number of PIC particles/cell etc.) correctly to avoid numerical heating. Typically, we take $64^3, 128^3, 256^3 $ grids (cells) {of} uniform cell size $\Delta x=\Delta y = \Delta z = 16$~a.u., time step $\Delta t = 0.1$ a.u., and approximately 15 particles/cell depending upon the cluster size.

\vspace{-0.25cm}
\subsubsection{PIC results and comparison with RSM}\label{subsubsec4}
\begin{figure}[]
	\centering
\includegraphics[width=0.45\textwidth]{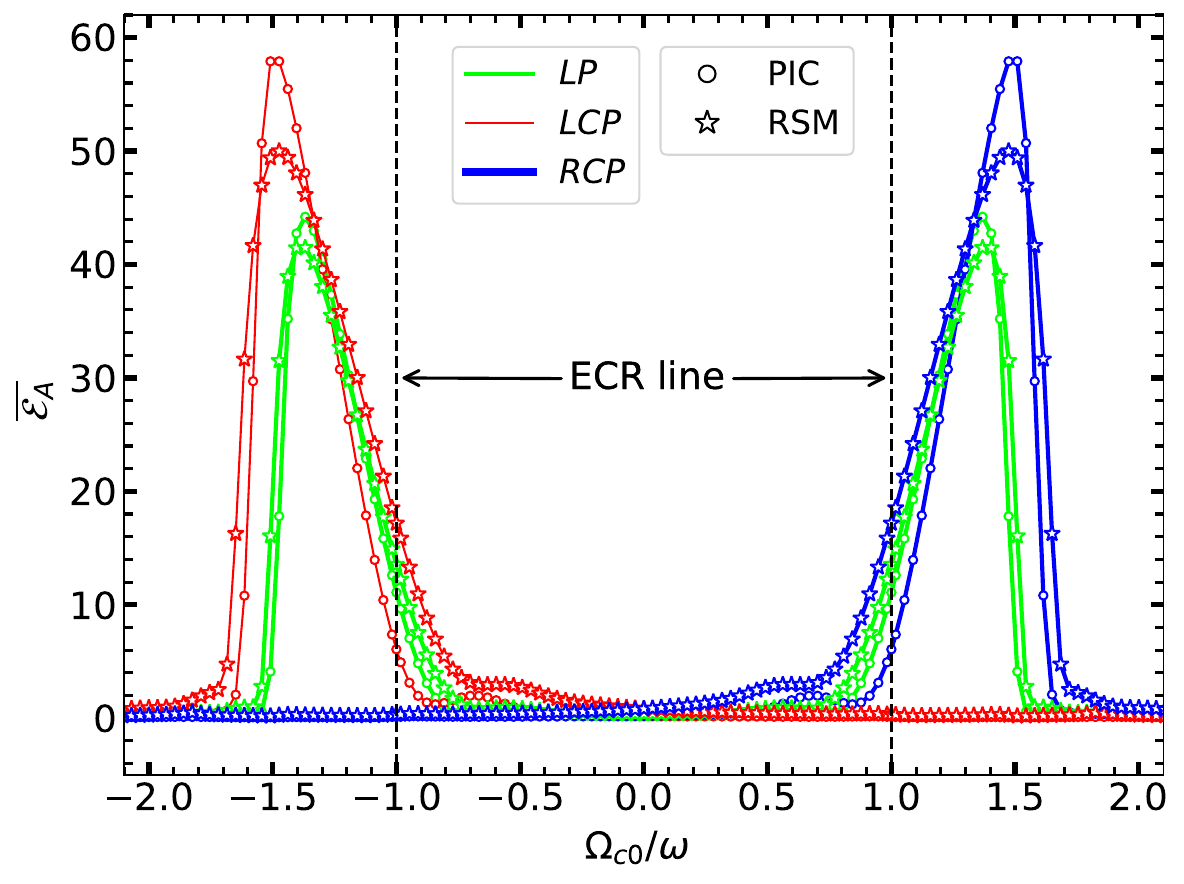}
 	\vspace{-0.25cm} 
 	\caption{Comparison of normalized average absorbed energy $\overline{\mathcal{E}_A}$ by a cluster electron for LP (green; light grey in grey scale) and CP laser field (blue and orange; black and dark grey in grey scale) in RSM ($\star$) and PIC ($\circ$) simulation within a range of $B_{ext}$ when $B_{ext} = \hat{z}B_0$ and $B_{ext} = -\hat{z}B_0$. The vertical dashed lines represents the ECR value of $B_0=0.057$~a.u.. Other parameters are same as in Fig.~\ref{Fig2}.}
 	%
 	%
 	\label{Fig5}
 \end{figure}

\vspace{-0.25cm}
In the previous works~\cite{Swain2022,Kalyani_PRA_2023}, we calculated the 
normalized absorbed energy $\overline{\mathcal{E}_A} = {\mathcal{E}_A}/N\Up$ per 
electron for a range of $B_0=0-2\omega$ with a LP laser field when 
$\vekt{B}_{ext} = + B_0\hat{z}$. Using both PIC and RSM, we showed that 
$\overline{\mathcal{E}_A}$ is not maximum at the ECR value of $B_0 = \omega$, 
instead the absorption peak occurred at a higher value of $B_0$ beyond the ECR 
due to the relativistically modified ECR (RECR) condition,
$\Omega_\mathrm{c} = \omegaC0/\gamma = \omega$. The same exercise is followed 
here for LP and CP (both RCP and LCP) laser fields with $B_{ext} = \pm 
B_0\hat{z}$, and $\overline{\mathcal{E}_A}$ is calculated for different $B_0$ values. Comparative results are shown in Figure~\ref{Fig5}. For 
the LP laser field with $B_{ext} = \pm B_0\hat{z}$, we find that $\overline{\mathcal{E}_A}$
is further enhanced from $\approx 13$ (at ECR) to the maximum $\approx 42$
for RSM (almost $3$ times) and $\approx 45$ for PIC at a shifted $B_0$
value from the ECR (the vertical dashed lines in Fig.~\ref{Fig5}). The 
absorption curve of $\overline{\mathcal{E}_A}$ show a
mirror symmetry for LP laser field across $B_{ext} = \pm B_0\hat{z}$. The 
single-electron RSM and the PIC results show a close matching.

In Fig.~\ref{Fig5}, we also show the RSM and PIC results for RCP and LCP laser fields with $B_{ext} = \pm B_0\hat{z}$ which are {\em central} to this work. Clearly, RCP with $B_{ext} = - B_0\hat{z}$ and LCP with $B_{ext} = + B_0\hat{z}$ leads to negligible $\overline{\mathcal{E}_A}$ in both RSM and PIC. Conversely,  RCP with $B_{ext} = + B_0\hat{z}$ and LCP with $B_{ext} = - B_0\hat{z}$ show maximum of $\overline{\mathcal{E}_A} \approx 50, 59$ for RSM and PIC respectively. Thus absorption per electron ${\mathcal{E}_A}/N$ are $\approx 8\Up, 14\Up$ higher compared to the respective LP cases (for RSM and PIC respectively). The absorption curves of $\overline{\mathcal{E}_A}$ vs $\Omega_{c0}/\omega$ with CP laser field also show a mirror symmetry in both PIC and RSM similar to the LP cases.

%
%


\vspace{-0.25cm}
\subsubsection{Angular distribution of PIC electrons}\label{subsubsec4}
\vspace{-0.25cm}
The energy and angular distributions of electrons are important for explaining their propagation as a beam. In an earlier report~\cite{Kalyani_PRA_2023}, we discussed the energy and angular distribution of PIC and RSM electrons in both position and momentum space for a LP laser field with $\vekt{B}_{ext} = +B_0\vekt{\hat{z}}$.
The energetic electrons showed a narrow cone-like propagation as a weakly
relativistic electron beam with an angular spread of $\Delta\theta<5^{\circ}$ at 
the ECR/RECR.
In Figure~\ref{Fig5}, it is shown that a CP laser field leads to higher amount of absorption $\sim 8 \Up, 14 \Up$ per electron (for RSM and PIC respectively) compared to the LP laser field.
In this section, we focus on the angular distribution of these PIC electrons with CP laser fields and compare with the LP cases corresponding to the results in Fig.\ref{Fig5}.

\begin{figure}[]
	\includegraphics[width=0.47\textwidth]{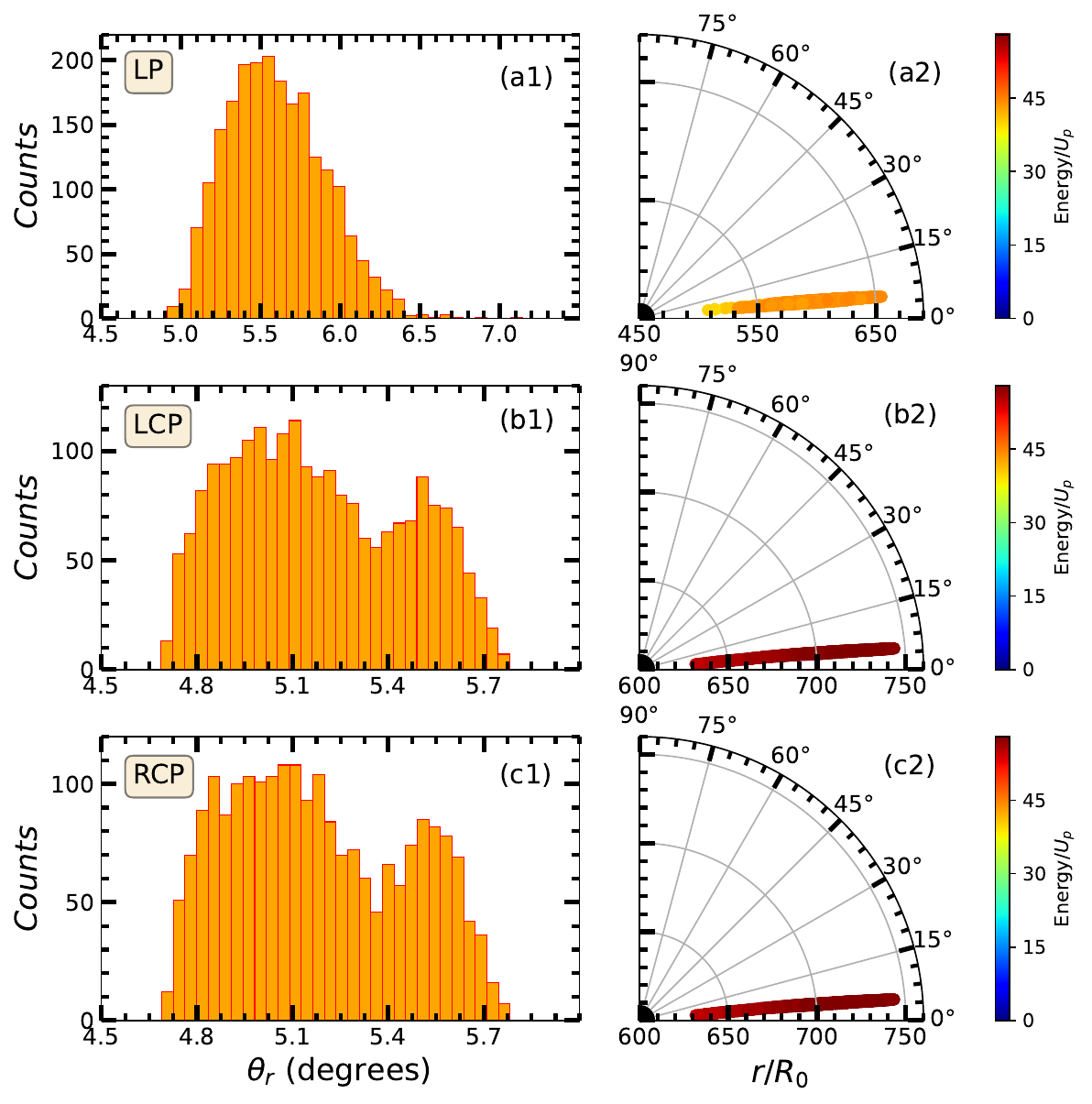}
	\vspace{-0.25cm} 
	\caption{Histograms showing the angular distribution of PIC electrons in the position space $\theta_r$ (left column) and respective polar plots with their normalized position $r/R_0$ vs $\theta_r$ (right column). These results correspond to the peak of the absorption curves in Fig.\ref{Fig5} with $B_0 = 0.079, 0.085$ a.u. for LP, CP respectively: LP with $B_{ext} = -\hat{z}B_0$ ($a1,a2$), LCP with $B_{ext} = -\hat{z}B_0$ ($b1,b2$), and RCP with $B_{ext} = \hat{z}B_0$ ($c1,c2$). The polar co-ordinates ($r,\theta_r$) are color-coded with their energy normalized by $U_p$. Other parameters are same as in Fig.~\ref{Fig5}.}
	%
	%
	\label{Fig6}
\end{figure}

{In the position space the angular deflection of an electron ($\theta_r$) is defined as, the angle made by the laser light while propagating along~$\vekt{z}$ (similar to the direction of $\vekt{B}_{ext} = B_0\vekt{\hat{z}}$) with its position vector $\vekt{r}$}. {This angle of elevation can be derived using}
\begin{equation}
	z=r\cos\theta_r, \,\, r_{\perp} = r\sin\theta_r; \,\,\, {\textrm{here}} \,\,\, r_\perp = \sqrt{x^2 + y^2}.
\end{equation}
In Figure~\ref{Fig6}, we plot the histogram of PIC electrons vs.~$\theta_r$ 
(Fig.~\ref{Fig6}a1-\ref{Fig6}c1, left column) and the respective polar plots 
(Fig.~\ref{Fig6}a2-\ref{Fig6}c2, right column) of their normalized position 
$r/R_0$ vs. $\theta_r$ corresponding to the peak points of 
$\overline{\mathcal{E}_A}$ curve in Figure~\ref{Fig5}. Polar co-ordinates 
($r,\theta_r$) are color-coded with the respective energy normalized by $\Up$. 
In the LP case, the electrons are spread over an angular range $\theta_r\approx 
5^{\circ}-7^{\circ}$ with the angular spread $\Delta \theta_r\sim 2^{\circ}$ 
(Fig.\ref{Fig6}a1) and the electron beam propagates to a distance $r\approx 660 
R_0$ (Fig.\ref{Fig6}a2). However, with LCP and RCP the angular spread  is 
further reduced to $\Delta \theta_r\sim 1^{\circ}$ within an angular range 
$\theta_r\approx 4.7^{\circ}-5.7^{\circ}$ (Fig.\ref{Fig6}b1,\ref{Fig6}c1) and 
the propagation distance is increased to $r\approx 750 R_0$ 
(Fig.\ref{Fig6}b2,\ref{Fig6}c2). In all three cases, the electron beam makes a very narrow 
cone-angle $\theta_r\approx 3^{\circ}-4^{\circ}$ with respect to the magnetic 
field direction $z$. This demonstrates that, with the CP laser field one may 
obtain energetically more intense electron beam of a very narrow angular spread 
$\Delta \theta_r \sim 1^{\circ}$ compared to LP laser field.

{The conical electron beam can also be elucidated by considering} the angular deflection in the momentum space ($\theta_p$) of the momentum vector $\vekt{p}$ w.r.t the laser propagation direction. {This angle, $\theta_p$, often holds significance in determining the transport characteristics of electrons and the type of magnetic configuration needed to facilitate their transport as a beam.}
With transverse momentum ($p_x,p_y$) and longitudinal momentum ($p_z$), we calculate $\theta_p$ from
%
\vspace{-0.5cm}
\begin{equation}
	p_z=p\cos\theta_p, \,\, p_{\perp} = p\sin\theta_p; \,\,\, {\textrm{where}} \,\,\, p_\perp = \sqrt{{p_x}^2 + {p_y}^2}. 
\end{equation}

\begin{figure}[]
	\includegraphics[width=0.47\textwidth]{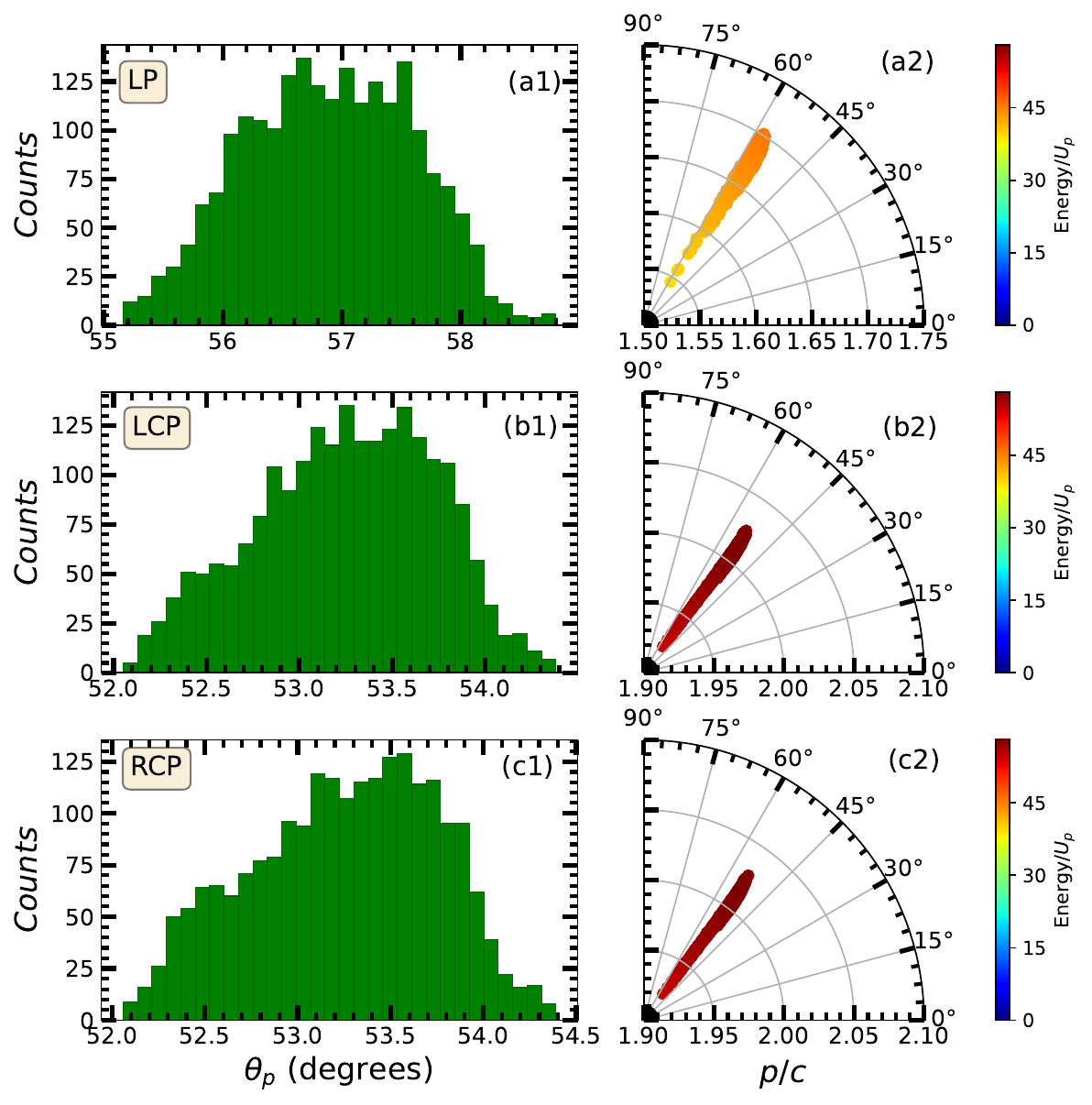}
	\vspace{-0.25cm} 
	\caption{
		Histograms showing the angular distribution of PIC electrons in the momentum space $\theta_p$ (left column) and respective polar plots  with their normalized momenta $p/c$ vs $\theta_p$ (right column). The panels ($a1,a2$),($b1,b2$), and ($c1,c2$) correspond to those panels ($a1,a2$), ($b1,b2$) and ($c1,c2$) in Fig.\ref{Fig6} respectively.The polar co-ordinates ($r,\theta_r$) are color-coded with their energy normalized by $U_p$. Other parameters are same as in Fig.~\ref{Fig5}.
		}
	%
	%
	\label{Fig7}
\end{figure}

In Figure~\ref{Fig7}, we plot the histogram of PIC electrons vs.~$\theta_p$
(Fig.\ref{Fig7}a1-\ref{Fig7}c1, left column) and the respective polar plots
(Fig.\ref{Fig7}a2-\ref{Fig7}c2, right column) of their normalized momentum $p/c$
vs.~$\theta_p$ corresponding to Fig.~\ref{Fig6} in the position space. Polar
co-ordinates ($p,\theta_p$) are color-coded with the respective energy 
normalized with $\Up$. In the LP case, the electrons are spread over an angular 
range $\theta_p\approx 55^{\circ}-59^{\circ}$ with an angular spread $\Delta 
\theta_p\sim 4^{\circ}$ (Fig.~\ref{Fig7}a1). The momenta of the beam electrons
reach weakly relativistic values $p\approx 1.7c$ (Fig.~\ref{Fig7}a2) in LP.
Again, with LCP and RCP the angular spread is reduced to $\Delta \theta_p \sim 
2^{\circ}$ within an angular range $\theta_p\approx 52^{\circ}-54^{\circ}$ 
(Fig.~\ref{Fig7}b1,\ref{Fig7}c1). The momentum of the beam electrons in these cases
reach higher values of $p\approx 2.03c$ (Fig.~\ref{Fig7}b2,\ref{Fig7}c2). In all three
cases, the electron beams make wide cone-angles of $\theta_p\approx
52^{\circ}-55^{\circ}$ and propagate like a spiral conical beam with respect to the
magnetic field direction $z$. This shows that, the CP case may produce 
energetically more intense electron beam of a very narrow angular spread $\Delta
\theta_p \sim 2^{\circ}$ of momentum compared to the LP case.


\section{Summary}
\label{sec6}
\vspace{-0.25cm}
The goal of this work is to study the laser energy absorption by cluster 
electrons in the collisionless regime for CP laser 
pulses (both LCP and RCP) when external magnetic fields $\vekt{B}_{ext} = \pm 
B_0\vekt{\hat{z}}$ and with LP laser pulses for $\vekt{B}_{ext} = - 
B_0\vekt{\hat{z}}$; which were not reported before. In earlier 
works~\cite{Swain2022,Kalyani_PRA_2023}, we have shown that the interaction of 
800~nm, 5-fs (FWHM) LP laser pulses for intensities $I_0 = 7\times 10^{16} - 
2\times 10^{17}\, \Wcmcm$ and $\vekt{B}_{ext} = + 
B_0\vekt{\hat{z}}$ with $B_0 = 0-2\omega$ a.u.,
enhances the average energy of a cluster electron 
${\mathcal{E}}_A\approx 36-70\,\Up$. This 
enhancement in ${\mathcal{E}}_A$ is a two stage 
process\cite{Swain2022} with anharmonic resonance (AHR) as the $1^{st} stage$ 
and electron cyclotron resonance (ECR) or relativistic ECR (RECR) as the $2^{nd} 
stage$. In the present work, we show that the energy enhancement is unaltered 
when $\vekt{B}_{ext} = - B_0\vekt{\hat{z}}$ with LP 
(Figs.\ref{Fig2} and \ref{Fig5}). But, ${\mathcal{E}}_A$ is more than $10 
\Up$ (Fig.\ref{Fig2} and Fig.\ref{Fig5}) higher with CP laser; particularly, when $\vekt{B}_{ext} = - B_0\vekt{\hat{z}}$ for LCP and $\vekt{B}_{ext} = + B_0\vekt{\hat{z}}$ for RCP 
(Fig.\ref{Fig5}). Otherwise, no further enhancement of 
${\mathcal{E}}_A$ occurs with $\vekt{B}_{ext} = \pm 
B_0\vekt{\hat{z}}$. Also, for both the LP and CP cases, the absorption curve 
with $\vekt{B}_{ext} = \pm B_0\vekt{\hat{z}}$ show a mirror symmetry.
The electrons propagate as a weakly relativistic electron beam 
(REB) in a conical-spiral having narrow opening angle of $6^\circ - 7^\circ$ 
in the position space (Fig.\ref{Fig6}) and wide opening angle 
($55^\circ-60^\circ$) in the momentum space (Fig.\ref{Fig7}) with respect to the 
laser propagation direction $\vect{z}$. The angular spread of electrons 
$\Delta\theta$ in both the position and the momentum space is $\approx 
4^\circ-5^\circ$ with LP \cite{Kalyani_PRA_2023}.
Present work shows that, the CP laser can further reduce the angular spread
$\Delta\theta \approx 2^\circ$ for both LCP and RCP with $\vekt{B}_{ext} = 
\pm B_0\vekt{\hat{z}}$ respectively. We have also performed PIC simulations for 
bigger clusters of deuterium, argon and xenon (not shown here for conciseness) 
with CP laser pulses under similar conditions of this work, which show great 
enhancement of total absorption beyond the LP 
cases \cite{Swain2022,Kalyani_PRA_2023} and formation of more intense electron beams
as shown in Figs.\ref{Fig6} and \ref{Fig7}.

Note that the
field components of a CP light does not vanish simultaneously. In addition, the 
phase 
matching condition (Fig.\ref{Fig4}) shows that one of the field components (or both) of CP
light may 
maintain the required phase with non-zero amplitude. This leads to higher 
absorption of the CP laser pulse than the LP case (Fig.\ref{Fig2} and 
Fig.\ref{Fig5}) in presence of an ambient
magnetic
field. It also improves the quality of the emitted electron beam in 
terms of divergence and energy in CP than LP (Fig.\ref{Fig6} and 
Fig.\ref{Fig7}).

This work may find importance for the fast 
ignition technique of inertial confinement fusion, laser-driven electron 
accelerators, and to understand energy absorption by plasma electrons as well 
as their dynamics under very strong field conditions in 
astrophysical environments, e.g., neutron stars and pulsars.

\vspace{-0.5cm} 
\section*{Acknowledgements}
\vspace{-0.25cm} 
Authors acknowledge Dr. Sagar Shekhar Mahalik for the initial help in numerical simulations and fruitful discussion; and Dr. Sudip Sengupta for careful reading of the manuscript.  
The numerical simulations presented in this work have been performed using Antya Linux cluster of HPC facility at IPR.

\bibliography{MSClusterKSMK}
\end{document}